\documentclass[prl,twocolumn,showpacs,amsmath,amssymb,floatfix]{revtex4-1}
\usepackage{epsfig}

\def\la{\langle}
\def\ra{\rangle}

\begin{document}

\title{Continuous phase amplification with a Sagnac interferometer}
\author{David J. Starling, P. Ben Dixon, Nathan S. Williams, Andrew N. Jordan and John C. Howell}
\affiliation{Department of Physics and Astronomy, University of Rochester, Rochester, NY 14627, USA }

\begin{abstract}
We describe a weak value inspired phase amplification technique in a Sagnac interferometer. We monitor the relative phase between two paths of a slightly misaligned interferometer by measuring the average position of a split-Gaussian mode in the dark port. Although we monitor only the dark port, we show that the signal varies linearly with phase and that we can obtain similar sensitivity to balanced homodyne detection. We derive the source of the amplification both with classical wave optics and as an inverse weak value.
\end{abstract}
\maketitle

\begin{figure}
\centerline{\includegraphics[scale=0.75]{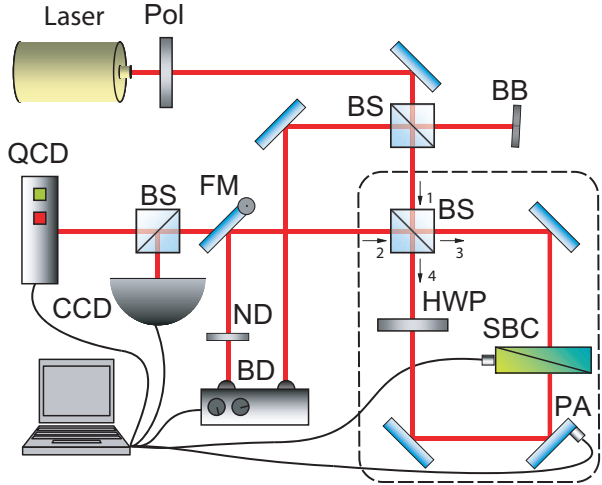}} \caption{(Color Online) Experimental setup. A coherent light source passes through a polarizer (Pol) producing horizontally polarized light before passing through the first 50/50 beamsplitter (BS). Half of the light strikes a beam block (BB) and is thrown out. The beam then enters the Sagnac interferometer via the second BS. One of the mirrors of the interferometer is controlled by a piezo actuator (PA) to precisely control the relative transverse momentum shift $k$. We use a half-wave plate (HWP) and a piezo-actuated Soleil-Babinet compensator (SBC) to produce a relative phase shift between the two light paths. During split-detection, we monitor the dark port using another BS that splits the light between a quad-cell detector (QCD) and a camera (CCD). During balanced homodyne detection, the interferometer is balanced and the flip-mirror (FM) is turned up, sending both bright and dark ports into the balance detector (BD). We use a neutral density filter (ND) in the bright port to correct for the light lost at the first BS.}
\label{exp}
\end{figure}

\textit{Introduction} ---
Phase measurements using coherent light sources continue to be of great interest in classical optics \cite{Sun1997,Erlandsson1988,Post1967,Schmitt1999,Wehr1999}. Not surprisingly, many advances in phase measurement techniques have been made since the introduction of the laser. For instance, Caves emphasized how the signal to noise ratio (SNR) of a phase measurement can be improved by using a squeezed vacuum state in the dark input port of an interferometer \cite{Caves1981}. Related advances in this area include the use of other non-classical states of light such as Fock states \cite{Holland1993} or the use of phase estimation techniques \cite{Higgins2009} which approach the Heisenberg limit in phase sensitivity \cite{Ou1997}. Unfortunately, these states of light tend to be weak and very sensitive to losses, in effect reducing the SNR of a phase measurement. As a result, the use of coherent light sources has dominated the field of precision metrology \cite{Abramovici1992,Dixon2009}. In this case, the phase sensitivity scales as $1/\sqrt{N}$ rad, where $N$ is the average number of photons used in the measurement.

Perhaps the most famous contemporary phase measurement device is the laser interferometer gravitational-wave observatory (LIGO) \cite{Abramovici1992,Abbott2009}. LIGO is a power-recycled Michelson interferometer with 4 km Fabry-Perot arms utilizing approximately ten Watts of input power. Although LIGO originally operated in a heterodyne arrangement using rf-sidebands, the second, ``enhanced" stage will make use of a homodyne configuration \cite{Waldman2006}. The expected phase sensitivity of this device will be around $10^{-14}$ rad. However, the low saturation intensity of even state-of-the-art detectors limits the number of photons one can use in a given measurement with a fixed bandwidth.

In this Letter, we show that it is possible to make a phase measurement with the same SNR as balanced homodyne detection yet only the light in the dark port is measured. We use a coherent light source with a split-detector in a Sagnac interferometer and show that the signal of a phase measurement is amplified. We derive our results using a classical wave description and summarize a quantum mechanical treatment which uses a similar weak value formalism to that presented in refs.\ \cite{Dixon2009,Starling2009}. Much like with weak values \cite{Aharonov1988,Duck1989}, there is a large attenuation (post-selection) of the electric field (number of photons). Thus we can, in principle, use a low-cost detector with a low saturation intensity and still obtain significantly higher phase sensitivity when compared to using a balanced homodyne detector with the same total incident intensity. This may play a crucial role in high-power phase measurement experiments such as LIGO.

\textit{Theory} ---
Consider a coherent light source with a Gaussian amplitude profile entering the input port of a Sagnac interferometer as shown in Fig.\ \ref{exp}. The interferometer is purposely misaligned using a piezo-actuated mirror such that the two paths experience opposite deflections. The transverse momentum shift imparted by the mirror is labeled as $k$. A relative phase shift $\phi$ can be induced between the two light paths (clockwise and counterclockwise) in the interferometer.

We model the electric field propogation using standard matrix methods in the paraxial approximation. We can then write the input electric field amplitude as 
\begin{equation}
\boldsymbol{E}_{in}=\left(
         \begin{array}{cc}
           E_0 e^{-x^2/4\sigma^2} & 0 \\
         \end{array}
       \right)^T,
\end{equation}
where $\sigma$ is defined as the Gaussian beam radius. The first position in the column vector denotes port 1 (see Fig.\ \ref{exp}) of the beam splitter and the second position denotes port 2 (with no input electric field). We assume that the beam is large enough so that the entire Rayleigh range fits within the interferometer. The matrix representation for the 50/50 beamsplitter is given by
\begin{equation}
\boldsymbol{B}=\frac{1}{\sqrt{2}}\left(
                      \begin{array}{cc}
                        1 & i \\
                        i & 1 \\
                      \end{array}
                    \right),
\end{equation}
where column and row one correspond to port 4 (counterclockwise) and column and row two correspond to port 3 (clockwise). We now define a matrix that gives both an opposite momentum shift $k$ and a relative phase shift $\phi$ between the two light paths
\begin{equation}
\boldsymbol{M}=\left(
\begin{array}{cc}
       e^{i(-k x+\phi/2)} & 0 \\
       0 &  e^{-i(-k x+\phi/2)}\\
\end{array}
\right).
\end{equation}
The exiting electric field amplitude is represented by the matrix combination $(\boldsymbol{BMB}) \boldsymbol{E}_{in}$,
\begin{equation}
\boldsymbol{E}_{out} = i E_0 e^{-x^2/4\sigma^2} 
	\left(
	\begin{array}{cc}
		-\sin(kx-\phi/2) \\
		\cos(kx-\phi/2)\\
	\end{array}
	\right),
\end{equation}
where the first position now corresponds to port 2 (the dark port) and the second position corresponds to port 1 (the bright port). 

For a balanced homodyne detection scheme, we take $k = 0$ and $\phi \rightarrow \pi/2 + \phi$ and subtract the integrated intensity of both ports. After normalizing by the total power, we obtain the unitless homodyne signal
\begin{equation}
\Delta_{h} = \sin(\phi).
\end{equation}
Thus, we see that by balancing the interferometer, we are measuring the signal along the linear part of the sine curve for small phase shifts. 

\begin{figure}
\centerline{\includegraphics[scale=0.9]{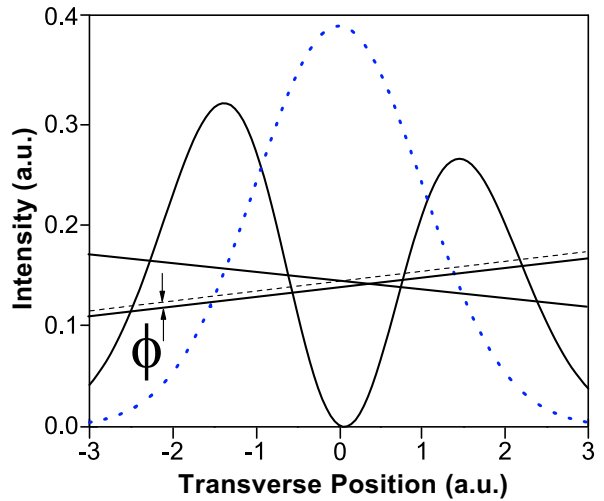}} \caption{(Color Online) Post-selected intensity distribution. The dotted (blue) curve is the single-mode input profile of the beam in the interferometer. The solid (black) curve is the post-selected split-Gaussian mode produced by the misalignment of the interferometer and a slight relative phase shift between the two light paths. The solid lines represent the tilted wave fronts in the two paths of the interferometer when they combine at the BS, one that is delayed relative to the other, producing an asymmetric split-Gaussian (shown). The dashed line represents a wave front with the same tilt but zero relative phase delay, which would result in a symmetric split-Gaussian (not shown).}
\label{twolobe}
\end{figure}

In contrast, if we consider a small transverse momentum shift ($k\sigma < 1$) and monitor only the dark port, given by the first element in the $E_{out}$ vector, we find that 
\begin{equation}
E^{(d)}_{out} \approx A\left(\frac{x}{\sigma}-\frac{\tan(\phi/2)}{k\sigma}\right)\exp[-x^2/4\sigma^2],
\end{equation}
where $A = -i E_0 k \sigma \cos(\phi/2)$. The intensity at the dark port is then given by
\begin{equation}
I_d(x) = P_{ps} I_0 \left(\frac{x}{\sigma}-\frac{\tan(\phi/2)}{k\sigma}\right)^2 \exp[-x^2/2\sigma^2],
\label{twolobeeq}
\end{equation}
where $P_{ps}$ is the attenuation (post-selection probability) of the measured output beam given by 
\begin{equation}
P_{ps} = |k\sigma\cos(\phi/2)|^2,
\end{equation}
and $I_0$ is the maximum input intensity density. Aside from the attenuation factor $P_{ps}$, $I_d(x)$ is normalized for vanishingly small $\phi$. Equation (\ref{twolobeeq}) is plotted in Fig. \ref{twolobe}.

In order to have the greatest accessibility for our results, we have presented the theory here in terms of classical wave optics. However, for an analysis at the single photon level, and an analysis of the signal-to-noise ratio for precision phase measurement, it is of interest to summarize the quantum mechanical derivation. The setup may be interpreted as the detection of the which-path (system) information of a single photon (clockwise or counter-clockwise). This is done indirectly using the transverse position degree of freedom of the photon as the meter, which is followed by a post-selection of the system state (due to the interference at the beam splitter), allowing only a few photons to arrive at the split-detector where the meter is measured.    

If the pre- and post-selected system states are almost orthogonal (so $k\sigma \ll \phi \ll 1$), then there is an anomalously large shift of the beam's position, referred to as the \textit{weak value} $A_w\approx -2i/\phi$ of the system (see Ref.\ \cite{Dixon2009}). The small overlap of the system's states gives rise to an amplification of the small momentum shift imparted by the mirror.  However, in the present case, $\phi \ll k\sigma \ll 1$, so the situation does not have a straightforward interpretation in terms of weak values.  Nevertheless, there are certain features in common:  the amplification effect can be traced back to the fact that there is a renormalization of the state, owing to a small post-selection probability $P_{ps}$. A quantitative analysis shows that the meter deflection is now proportional to the \textit{inverse weak value} $A_w^{-1}$, 
\begin{equation}
\la x\ra = -2\, {\rm Im} A_w^{-1} \approx - \phi/k,
\label{shift}
\end{equation}
which may be interpreted as an amplification of the small phase shift by the mirror's momentum shift $k$. Notice that in contrast to the usual weak value case, the post-selected distribution is not a simple shift of the meter wavefunction, but in fact corresponds to a two-lobe structure [Eq.\ (\ref{twolobeeq})] as seen in Fig.\ \ref{twolobe}.

Instead of measuring the average position, one can use a split-detection method by integrating the intensity over the right side of the origin and subtracting from that the integrated intensity over the left side of the origin. This detection method is well suited to the split-Gaussian beam and results in a split-detection signal which, if normalized by the total power striking the detector, is proportional to the average position. This quantity is given by 
\begin{equation}
\Delta_s \approx -\sqrt{\frac{2}{\pi}} \frac{\phi}{k \sigma} \approx \sqrt{\frac{2}{\pi\sigma^2}} \la x \ra.
\label{signal}
\end{equation}

Despite the large amplification of the average position measurement of the post-selected photons, the SNR is essentially the same for a balanced homodyne measurement of phase. The SNR of a phase measurement using balanced homodyne or split-detection can be expressed as $\mathcal{R}_{h,s} = \Delta_{h,s}\,\sqrt{N_d}$, where $N_d$ is the number of photons striking the detector. These expressions are identical, except for an overall constant factor of $\sqrt{2/\pi}$. This reduces the SNR of the split-detection method by approximately 20\%. It is also interesting to note that the SNR is independent of $k$. Thus, we can in principle reduce $k$ (and $P_{ps}$) arbitrarily, allowing us to increase the input power and therefore $N$, ultimately improving the measurement sensitivity arbitrarily while using the same detector. 

\textit{Experiment} ---
In the present experiment (see Fig.\ \ref{exp}), the coherent light beam was created using an external cavity diode laser tuned near the $D_1$ line of rubidium, approximately 795 nm. The beam was coupled into single mode fiber and then launched to produce a single mode Gaussian profile. The light was collimated with a radius of $\sigma \approx 775 \,\mu\mbox{m}$ and the continuous wave power ranged from 0.5 mW up to 1 mW. The Sagnac, composed of a 50/50 beam splitter and three mirrors, was rectangular. We used two configurations for the geometry of the interferometer, one with dimensions 39 cm x 8 cm (large) and another with dimensions 11 cm x 8 cm (small). The beam profile and position of the post-selected photons were measured using a quad-cell detector (QCD, New Focus model 2921) functioning as a split-detector and a CCD camera (Newport model LBP-2-USB). During balanced homodyne detection, the signal was measured using a Nirvana balance detector (BD, New Focus model 2007). The quantum efficiency of the BD was about 81\%, whereas the quantum efficiency of the QCD was 75\%. The QCD was also equiped with a protective glass plate with only 50\% transmissivity. The outputs from the QCD and the BD were fed into two low-noise preamplifiers with frequency filters (Stanford Research Systems model SR560) in series.

\begin{figure}
\centerline{\includegraphics[scale=0.78]{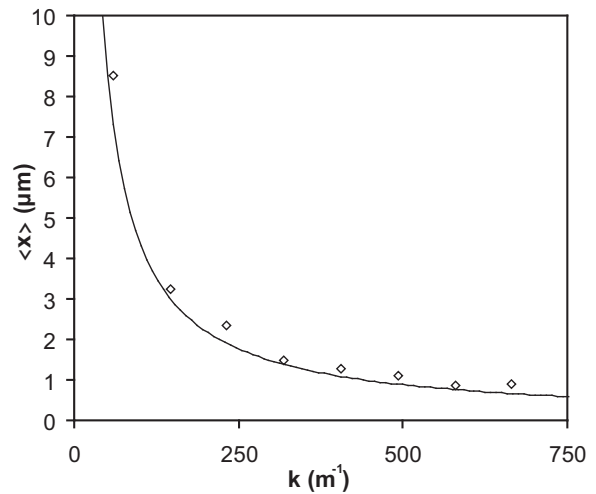}} \caption{Dependence on transverse momentum. The transverse momentum shift imparted by the piezo-actuated mirror was varied and the split-detection signal was measured using the QCD. The solid line is the theory curve from Eq.\ (\ref{shift}) using the expected phase shift.}
\label{dk}
\end{figure}

\begin{figure}
\centerline{\includegraphics[scale=0.85]{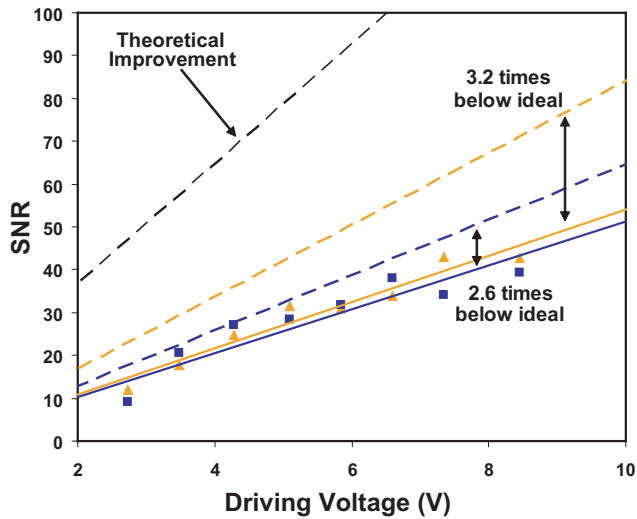}} \caption{(Color Online) Experimental comparison of split-detection to balanced homodyne. We vary the driving voltage applied to the piezo-actuator and measure the SNR using balanced homodyne detection (orange triangles) and the split-detection method (blue squares). The input power to the interferometer is approximately the same for both methods. Linear fits to the data (solid lines) show that these two methods have essentially the same sensitivity. The ideal quantum limited SNR (factoring in the quantum efficiency of each detector) is plotted using a dashed blue line (split-detection) or a dashed orange line (balanced homodyne detection). The dashed black line illustrates the theoretical ($\sqrt{N}$) improvement of the split-detection method assuming that an equal number of photons are incident on both the split-detector and the balance detector.}
\label{homo}
\end{figure}

We used a half-wave plate (HWP) with a piezo-actuated Soleil-Babinet Compensator (SBC) inside the Sagnac interferometer to induce a relative phase shift. The HWP was oriented such that the horizontally polarized input light was rotated to vertically polarized light. The SBC was oriented such that the fast axis was vertical and the slow axis was horizontal. The two light paths in the interferometer encountered these optical elements in opposite order, allowing for a known, tunable phase difference between them. The piezo-actuator in the SBC, which moved approximately 100 pm/mV, imparted a relative phase shift of $22\pm0.9\,\mu$rad/V.

Using the large configuration, with 0.5 mW of input power, the piezo actuator in the SBC was driven with a 20 V peak to peak sine wave at 634 Hz, corresponding to a relative phase shift of 440 $\mu$rad. The normalized split-detection signal $\Delta_s$ (factoring in an offset from junk-light hitting the detector) was measured while the transverse momentum shift $k$ was varied using the piezo-actuated mirror. After scaling $\Delta_s$ by the appropriate factor given in Eq.\ (\ref{signal}), the results were plotted in Fig.\ \ref{dk}. The theory line, which corresponds to a relative phase shift of 440 $\mu$rad, is drawn along with the data. We see good agreement of the data with theory, with a clear inverse dependence of $\la x \ra$ on $k$. However, it should be noted that an determination of $k$ for this fit requires calibration, which in practice is quite simple.

We then compared this split-detection method of phase measurement to a balanced homodyne measurement. We used the small configuration with 625 $\mu$W of (effective) continuous wave input power---taking into account various attenuators---and varied the driving voltage to the piezo-actuator in the SBC. The low-pass filter limits the laser noise to the 10\% to 90\% rise-time of a 1 kHz sine wave ($300\mu$s). We take this limit as our integration time to determine the number of 795 nm photons used in each measurement. We measured the SNR of a phase measurement (see Fig.\ \ref{homo}) using the same method as ref. \cite{Starling2009} and found that the SNR of our homodyne measurement was on average 3.2 below an ideal quantum limited system. The SNR of our split-detection method was on average 2.6 times below an ideal quantum limited system. We take into account the quantum efficiency of each detector for these two values, yet we ignore any contribution of dark current to the expected SNR. 

Importantly, the SNR resulting from both measurement techniques is approximately the same. However, the split-detection method for this data had only about 15\% of the input light incident on the detector. Thus, for diodes with the same saturation intensity, it is possible to use almost seven times more input power with this configuration, resulting in a SNR about 2.6 times higher (the black, dashed line in Fig.\ \ref{homo}). The improvement of the SNR by taking advantage of the attenuation before the detector has no theoretical limit and is only limited in practice by phase front distortions and back-reflections off of optical surfaces which degrade the fidelity of the interference. Using commercially available equipment and one day of integration time, sub-picoradian sensitivity is possible even with a low-saturation intensity split-detector.

\textit{Conclusion} ---
In summary, we have shown that the measurement of a relative phase shift between two paths in an interferometer can be measured and amplified using a split-detection method. Furthermore, this method is comparable to the sensitivity achievable using balanced homodyne techniques, yet only one output port of the interferometer is measured. In addition, the split-detector can have a low saturation intensity due to the large post-selection attenuation. In fact, the higher the attenuation, the larger the amplification of the split-detection signal. Furthermore, although we have described this experiment classically, we have shown that this technique exhibits an inverse weak value. As discussed in refs.\ \cite{Hosten2008,Starling2009}, these ``weak value" type experiments have the added benefit of reducing technical noise.

This phase detection method has applications in a number of fields, e.g.\ magnetometry (using nonlinear magneto-optical rotation) or rotation sensing. We believe that this technique is a robust, low-cost alternative to the balanced homodyne phase detection method. 

This work was supported by the Army Research Office, DARPA DSO Slow Light and a DOD PECASE award.


\end{document}